\documentclass[conference]{IEEEtran}
\IEEEoverridecommandlockouts
\usepackage{cite}
\usepackage{amsmath,amssymb,amsfonts}
\usepackage{algorithmic}
\usepackage{graphicx}
\usepackage{textcomp}
\usepackage{xcolor}
\usepackage[ruled,vlined]{algorithm2e}
\def\BibTeX{{\rm B\kern-.05em{\sc i\kern-.025em b}\kern-.08em
    T\kern-.1667em\lower.7ex\hbox{E}\kern-.125emX}}
\begin{document}

\title{Crowdsourcing Controller - Utilizing Reliable Agents in a Multiplayer Game
\thanks{This paper is partially supported by the EU Horizon 2020 research and innovation programme under the MSCA grant No. 893667}
}


\author{\IEEEauthorblockN{1\textsuperscript{st} Kacper Kenji Lesniak}
\IEEEauthorblockA{
\textit{University of Copenhagen and Dynasty Studios}\\
Copenhagen, Denmark \\
kkl@di.ku.dk}
\and
\IEEEauthorblockN{2\textsuperscript{nd} Maria Maistro}
\IEEEauthorblockA{
\textit{University of Copenhagen}\\
Copenhagen, Denmark \\
mm@di.ku.dk}
\and
}

\maketitle

\begin{abstract}
This paper presents a new use case for continuous crowdsourcing, where multiple players collectively control a single character in a video game. 
Similar approaches have already been proposed, but they suffer from certain limitations: (1) they simply consider static time frames to group real-time inputs from multiple players; (2) then they aggregate inputs with simple majority vote, i.e., each player is uniformly weighted.
 
We present a continuous crowdsourcing multiplayer game equipped with our Crowdsourcing Controller.
The Crowdsourcing Controller addresses the above-mentioned limitations: (1) our Dynamic Input Frame approach groups incoming players' input in real-time by dynamically adjusting the frame length; (2) our Continuous Reliability System estimates players' skills by assigning them a reliability score, which is later used in a weighted majority vote to aggregate the final output command.

We evaluated our Crowdsourcing Controller offline with simulated players and online with real players.
Offline and online experiments show that both components of our Crowdsourcing Controller lead to higher game scores, i.e., longer playing time.
Moreover, the Crowdsourcing Controller is able to correctly estimate and update players' reliability scores.

\end{abstract}

\begin{IEEEkeywords}
Crowdsourcing, Multiplayer, Reliability system
\end{IEEEkeywords}

\section{Introduction}

Crowdsourcing systems are fascinating and powerful tools that can be utilized in a variety of use cases~\cite{brabham-crowdsourcing}. Thanks to wide access to a stable and fast network connection, it is possible now to develop those systems for real-time applications, i.e., continuous crowdsourcing. In recent years, research has been dedicated to real-time and online crowdsourcing systems, with use cases such as video captioning, robot controls, audio processing, and assistive technology~\cite{real-time-crowd-control}.

Another topic that emerged in recent years are video game streams, where spectators become active players by voting on certain gameplay choices or even taking direct control of the game. Recent research has found that watching a stream of a video game is more about social interaction than the game content itself~\cite{watching-games}. This type of games, where larger crowds play and cooperate together through aggregated network input are described as Emergent Multiplayer Games (EMG)~\cite{emergent}. The popularity of these events shows that there is a market for new, innovative ways in which players can cooperate in game contexts. Although research has been put into how a multiplayer voting mechanisms can work from a gameplay perspective~\cite{emergent}, there is little insight into merging crowdsourcing approaches with multiplayer video game controllers.

In this paper, 
we explore continuous crowdsourcing approaches in the context of multiplayer games, in which players make collective decisions regarding the gameplay.
We present a new use case for continuous crowdsourcing, where multiple players collectively control a single character in a video game. 
Similar approaches have already been proposed in different contexts, but they suffer from certain limitations.
They consider static frames to group inputs from multiple users 
and they do not utilize approaches commonly implemented in crowdsourcing systems to estimate players reliabilities when aggregating inputs.


The contribution of this paper is twofold. The first part is a simple video game with an innovative mechanic that lets multiple network players control one character through an aggregated input.
The second part is our Crowdsourcing Controller, which controls how to: (1) group real-time inputs coming from multiple users into frames; (2) aggregate inputs within a frame and issue a single output command.
We propose two novel approaches to aggregate and process crowdsourced inputs: the Dynamic Input Frame to group inputs and the Continuous Reliability System to aggregate them. 
The former aims at detecting a precise moment in time when the input should be issued to the game, while the latter focuses on recognizing and utilizing reliable players during the play session.
A combination of offline, AI-based experiments, as well as online experiments, with $9$ real players, has been conducted to gather the data needed for the analysis and evaluation of the proposed solutions. 



The paper is organized as follows: related work on crowdsourcing and online multiplayer games (\S~\ref{sec:related}); the online game (\S~\ref{chap:system}); our Crowdsourcing Controller (\S~\ref{chap:system}); experimental evaluation (\S~\ref{sec:evaluation-em}); conclusions and future work (\S~\ref{subsec:conclusion}).

\section{Related Work}
\label{sec:related}

First, we describe crowdsourcing, its definitions and most prevalent use cases  
(\S~\ref{subsec:crowdsourcing}).
Then we present related work on Emergent Multiplayer Games (EMG), a type of online video streams in which viewers have direct control over the game through voting mechanisms (\S~\ref{subsec:emg}).

\subsection{Crowdsourcing} 
\label{subsec:crowdsourcing}

The term crowdsourcing firstly appeared in $2006$ with the meaning of ``taking a function once performed by employees and outsourcing it to an undefined (and generally large) network of people in the form of an open call''~\cite{howe_2006}.
Over the years, this idea has gained much interest both in industry and academia and became an umbrella term that encompasses many different use cases, as distributed-human-intelligence tasks, knowledge discovery and management, etc.~\cite{brabham-crowdsourcing}.
One of the most famous examples is Amazon Mechanical Turk, a crowdsourcing platforms where requesters can submit small jobs, such as image labeling, called Human Intelligence Tasks (HITs), which are performed by a remote community of workers.
In recent years, researchers explored novel ideas to use crowdsourcing such as software development~\cite{software-dev} and even writing fiction~\cite{novel}.

A persistent problem in crowdsourcing is how to engage crowd workers when performing HITs.
Usually crowd workers are paid a small amount of money~\cite{SilbermanEtAl2018}, which does not guarantee that workers perform to the best of their ability, even if a fair payment can be helpful~\cite{dEonEtAl2019,YeEtAl2017}.
Another way of engaging workers is through gamification: in~\cite{crowdsourcing-game} a novel platform is designed, where users playing an online game were labeling images while competing against each other.




Providing engaged and motivated workers does not always guarantee that their answers will be correct. 
Therefore each HIT is usually assigned to multiple crowd workers, whose inputs are aggregated with statistical methods.
The simplest way to aggregate multiple answers is by Majority Vote (MV), i.e., the answer that is the most popular among the workers is predicted to be correct. 
This, however, does not take into consideration differences in workers' expertise. 


Expectation Maximization (EM) algorithm has been used as early as in~\cite{Dawid79maximumlikelihood} to tackle this problem. 
EM iteratively estimates the probability that a worker correctly completes a set of HITs (this probability is called reliability in \S~\ref{chap:system}).
Some limitations of EM are that it needs to run offline on the whole dataset and it is costly due to its iterative nature. 
Both of these arguments prevent EM application in real-time scenarios.

A relatively new type of crowdsourcing is \textit{Continuous crowdsourcing}~\cite{real-time-crowd-control} which engages workers over long periods and allows them to maintain the context of the task.
Moreover the tasks change over time reacting to workers' input.
The responses from the workers are processed in so-called \textit{input mediators} that aggregate them to produce the system output.
Continuous crowdsourcing has been applied to a variety of use cases, for example, real-time continuous video/audio captioning to collect activity recognition labels from video stream~\cite{interactive-crowds}, and a robot control system able to react to unforeseen circumstances\cite{real-time-crowd-control}.

Continuous crowdsourcing requires to dynamically estimate workers' reliabilities. 
Such calculations need to be done in real-time, as workers' input is immediately applied to the task at hand and different reliabilities affect the votes aggregation. 
In~\cite{online-crowdsourcing} a modified version of the EM algorithm is used in real-time to estimate workers' reliabilities. 
The authors, however, do not mention the time complexity of this approach, and using an iterative method in real-time scenarios might not be optimal. In this paper, we propose a method of calculating workers reliabilities in a real-time scenario without the need to perform high complexity calculations.

\subsection{Video Games Controlled by Multiple Players}
\label{subsec:emg}


There is a growing trend of multiplayer video games, where multiple network users control a single instance of a video game streamed through an online platform.
One of the first, and to this day the most popular event using this approach was Twitch Plays Pokemon, where thousands of players watched a video stream of an ongoing Pokemon Red game while issuing commands through the chat.
The game creators defined two modes: \textit{anarchy}, all inputs from all players are forwarded to the game; and \textit{democracy}, all inputs are considered as votes, counted and then forwarded to the game, similarly to input mediators 
(\S~\ref{subsec:crowdsourcing}).
The democracy mode resulted in a more conservative and slower behavior of the character, but allowed the players to beat the hardest challenges.

Twitch Plays Pokemon is one of the earliest examples of an Emergent Multiplayer Game (EMG)~\cite{emergent}, a video game in which large crowds play collectively through a network voting system.
EMGs utilize a set of relatively simple game mechanics and subsystems, that when controlled by large crowds result in creative and unique ways of playing.
They encourage coordination and self-organization of the crowd, both in-game and outside of it. 
Explored mechanics usually use a voting system that gathers input from a large crowd and applies the results after a certain delay. 
The time reserved for collecting the input, however, makes the system not adequate for fast-paced, real-time scenarios. We address this challenge using our Crowdsourcing Controller.

\section{System Overview and Use Case}

\begin{figure}[tb]
  \centering
  \includegraphics[width=0.9\columnwidth]{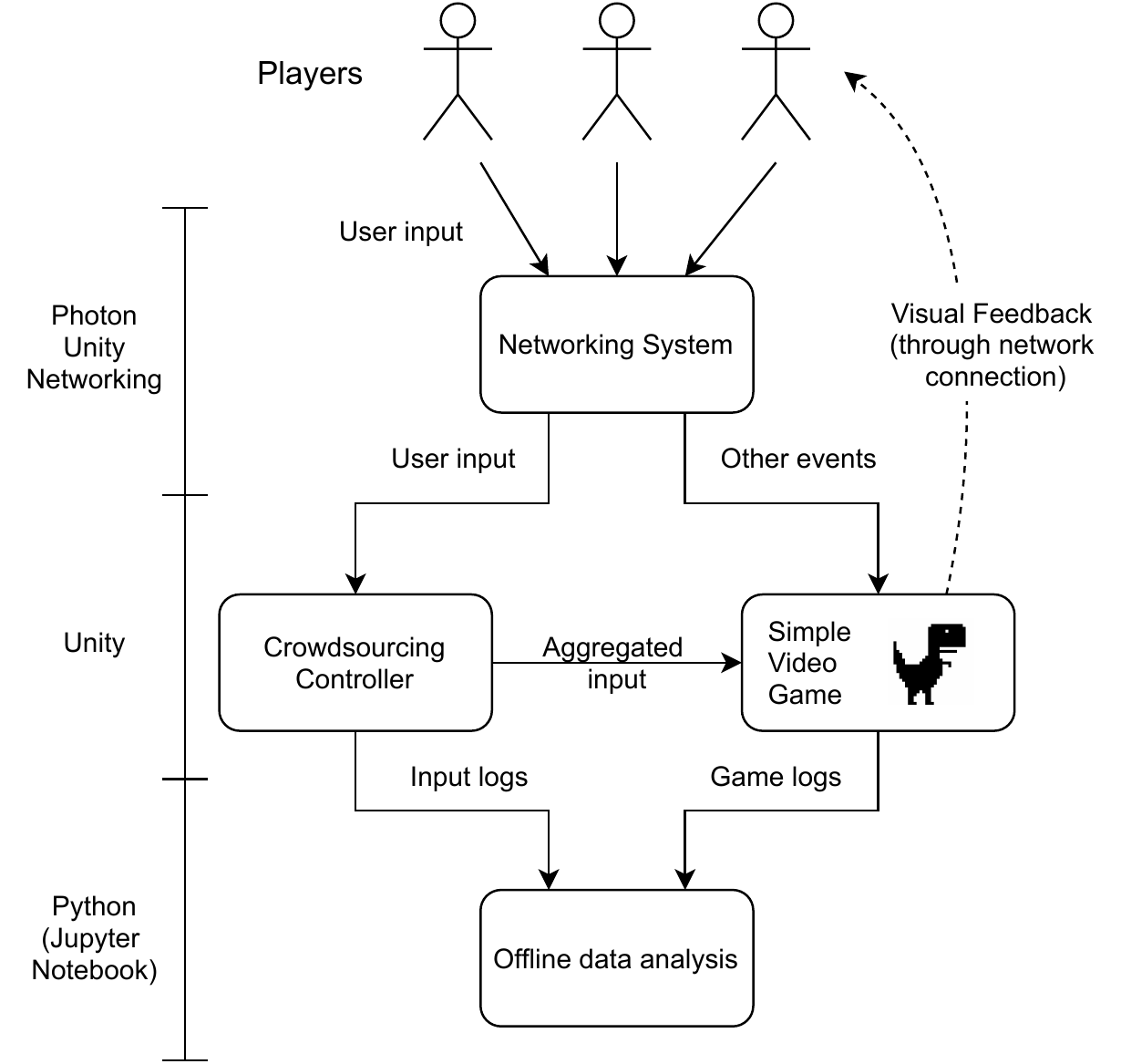}
  \caption{General system overview with technologies used for implementation}
    \label{fig:system-overview}
\end{figure}

Our continuous crowdsourcing system is presented in Figure~\ref{fig:system-overview}.
The central part of the system is a simple video game (described in \S~\ref{subsec:video_game}). 
Players issue inputs through the network as if they were the only player in control of the game character. 
The input is aggregated and processed by the \textit{Crowdsourcing Controller}. 
The output of the system is a singular game command, whose type and issue time are calculated based on all the inputs sent over the network. 
The aggregated game command is applied with immediate effect, which influences the state of the game and is seen by all players on their screens. 
The source code of the implementation of our system is publicly available\footnote{https://github.com/KacperKenjiLesniak/crowded-dinosaur}.

\subsection{Use case: a Simple Video Game}
\label{subsec:video_game}

Our Crowsourcing Controller controls a character in a simple video game which we describe next.
We choose this use case because: (1) crowd workers are generally familiar with simple controls as well as typical challenges presented in video games; (2) we wanted to explore the presented crowdsourcing system as part of a novel gameplay mechanic; (3) gamification has proven to be a viable method to engage users in a variety of tasks~\cite{gamification} which can help with keeping workers motivated throughout the course of experiments.

\begin{figure}[tb]
  \centering
  \includegraphics[width=0.9\columnwidth]{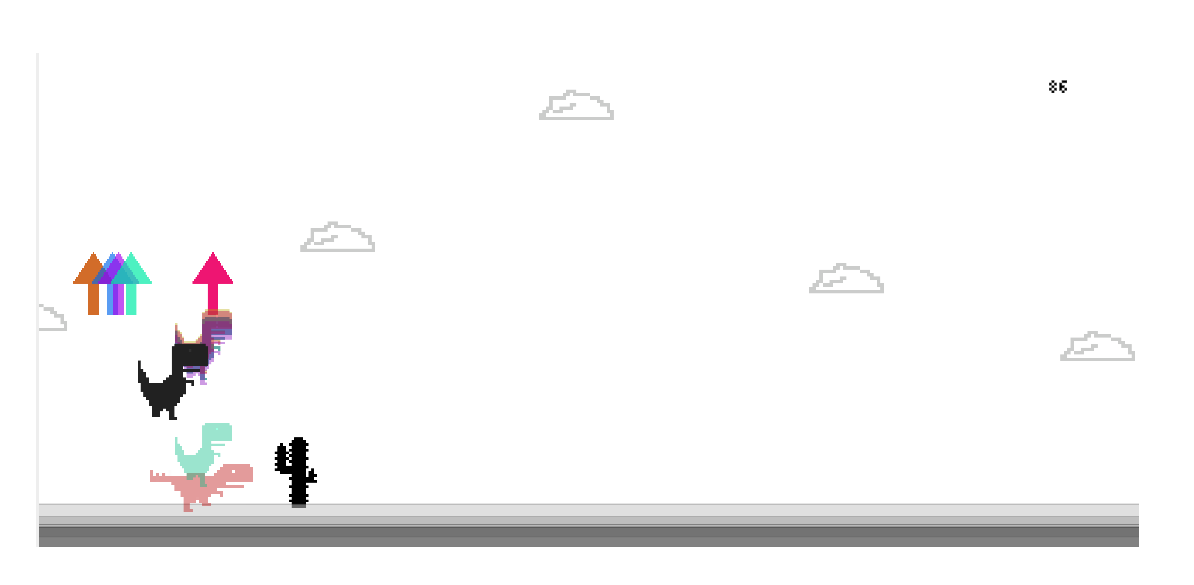}
  \caption{Crowd controlled dinosaur (black) and player dinosaurs (different colors) jumping over an obstacle}
    \label{fig:game2}
\end{figure}

The implemented game is a clone of a well-known mini-game played in Chrome (see Figure~\ref{fig:game2}).
Players are assigned their own dinosaur character of a different color.
The crowd-controlled dinosaur which corresponds to the aggregated output is represented by a black dinosaur.
Players' dinosaurs behave like ``ghosts'' and they serve purely as a visual clue to the players (Figure \ref{fig:game2}).
Thanks to them the players can see that their input is being processed and is working as intended, even when the crowdsourced dinosaur is not reflecting their actions at the current moment.
The players also see each other characters, which can enhance engagement and competition, and it provides further visual feedback.
Players seeing the black dinosaur impacting an obstacle, even when they issued a correct input, can feel frustrated and believe that the system is not working. 
Seeing that others failed to issue the correct input can help alleviate this issue. Moreover, seeing all other characters can lead to players quickly realizing that they are issuing the inputs out of sync. 

We chose this game because it meets a specific set of requirements: the input comes in real-time and jumping or crouching at the right moment is crucial for progression; the input is discrete (short jump, long jump, crouch); there is a player score which can be used as an estimate of player skills, and the game is widely known and simple to play.

\section{Crowdsourcing Controller}
\label{chap:system}

We describe the Crowdsourcing Controller next, which is responsible for analyzing the inputs coming from the players, aggregating them, and issuing a single command that best represents the combined inputs.

\begin{figure}[h]
  \centering
  \includegraphics[width=0.9\columnwidth]{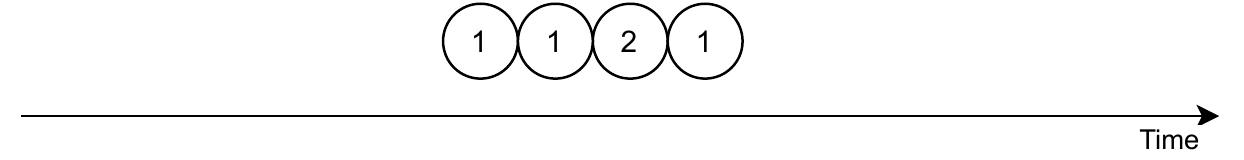}
  \caption{Inputs of $2$ types issued by $4$ players in a short period of time}
    \label{fig:inputs}
\end{figure}

Let us consider a system where four players are able to issue two types of input - $1$ and $2$. 
The example in Figure~\ref{fig:inputs} presents a certain moment in time, during which the system receives four inputs, each coming from a different player, three of them are of type $1$ and one is of type $2$. 

This situation poses two challenges. 
The first challenge relates to the definition of the \textit{input frame}, i.e., the time window where all inputs correspond to the players' reaction to the same game event, for example an obstacle appearing on screen.
In the example in Figure~\ref{fig:inputs}, all $4$ inputs  are issued at a similar time, so it is reasonable to consider them in the same group or input frame.
As for when the input frame should be analyzed further by the system, in the presented example it is reasonable to do it immediately after receiving the last input. 
Moreover, since no more inputs are issued after the last one, all inputs can be grouped and processed by the system for further analysis.
However, it is not possible to make similar decisions in a real-time setting without knowing what comes next.
Section~\ref{subsec:frames} describes how we tackle this problem with dynamic and static input frames.

The second challenge relates to the aggregation approach, i.e., how to combine multiple inputs in a single one. 
In the presented example, three players voted for input $1$ and one player for input $2$, so according to majority vote, the output of the system should be $1$. 
However, majority vote treats all payers equally and does not consider the expertise of each player, e.g., the vote of an expert or reliable player can weight more than the vote from a beginner and lead to a more accurate aggregated score. 
Section~\ref{subsec:input_aggregation} describes how we estimate such weights, called reliabilities.



\begin{figure}[htb]
  \centering
  \includegraphics[width=0.9\columnwidth]{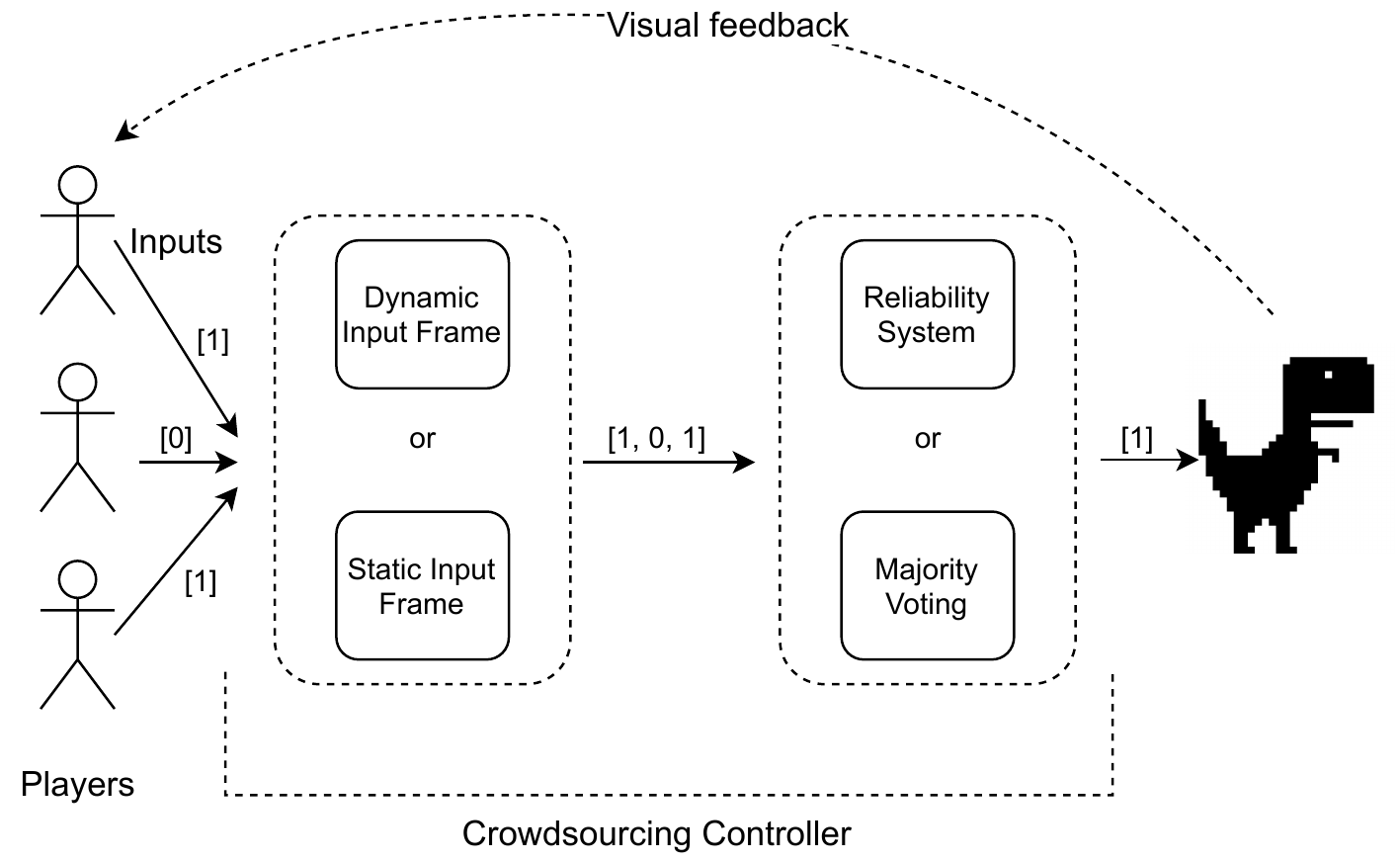}
  \caption{Flow of players' inputs through different input mediators}
    \label{fig:system-breakdown}
\end{figure}

The above mentioned challenges are handled by \textit{input mediators}, illustrated in Figure~\ref{fig:system-breakdown}. 
Input mediators are components of the crowdsourcing system that: (1) process the input stream coming in real-time from several players and group inputs into frames; and (2) produce the final aggregated output that is issued to the game.
These two components together define our Crowdsourcing Controller.

\subsection{Dynamic and static Input Frames}
\label{subsec:frames}

The first component of the Crowdsourcing Controller groups continuous input into input frames, which contain no more than a single input for each player (a missing input from a player is represented as $0$). 
A frame represents a short moment in time in which the users issue the commands which can result in a specific behavior of the system.
%
We use $2$ types of input frames, static and dynamic, which we describe next.

The Static Input Frames~\cite{real-time-crowd-control} slice the input stream into frames of a constant length.
In case of multiple inputs from the same player in a single frame, the system uses the most recent one. 
One limitation of this approach is how to define the frame length: too short frames will wrongly group inputs related to the same event in different frames, too long frames will negatively affect the system responsiveness.

\begin{figure}[tb]
  \centering
  \includegraphics[width=0.9\columnwidth]{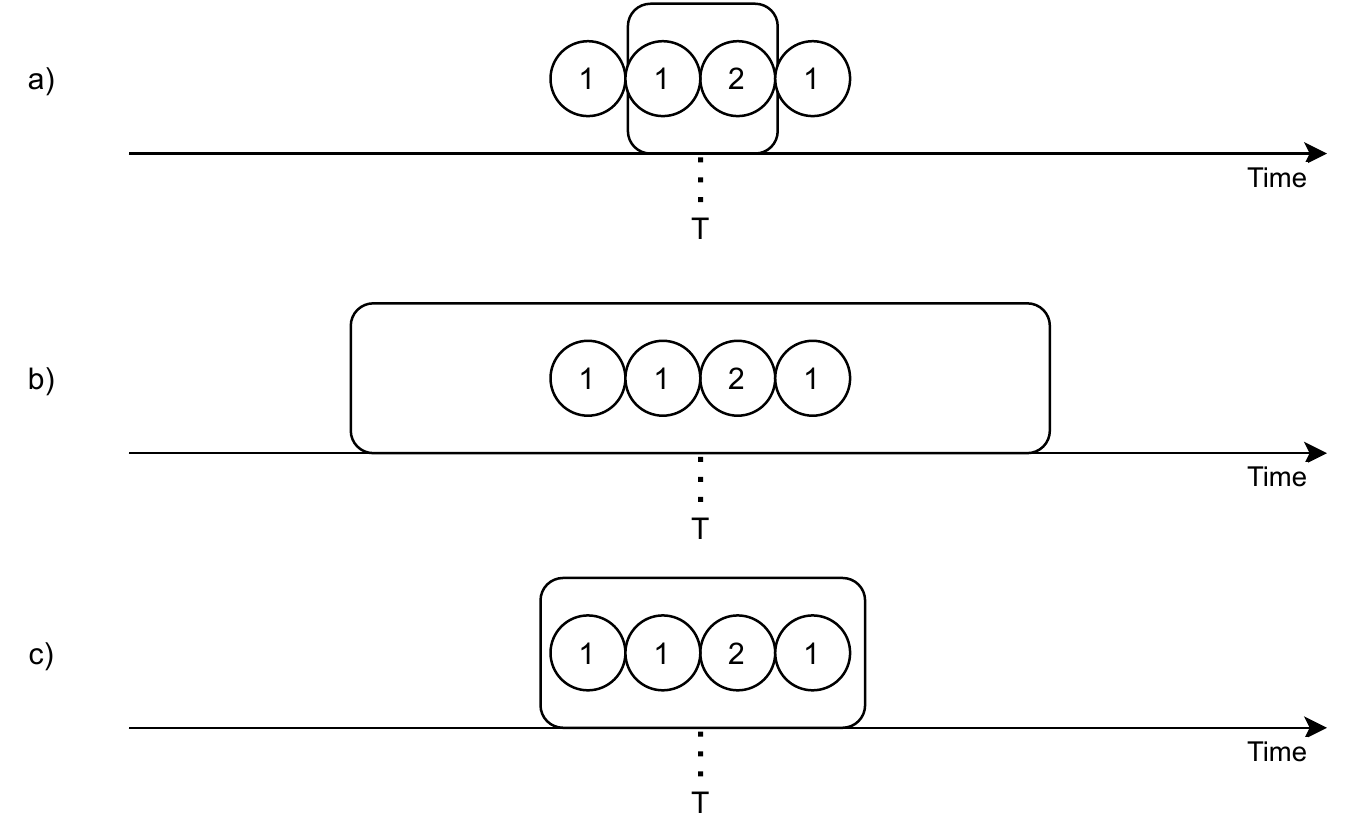}
  \caption{Dynamic frames, threshold (T) of 2 players out of 4 marks the center of the frame: a) too short duration, part of input ignored; b) too long duration, unresponsive system; c) correct duration for the situation}
    \label{fig:frames-dynamic}
\end{figure}

To address this limitation, we propose the Dynamic Input Frame approach, where the frame length is dynamically defined based on players' inputs.
The system forms a queue of incoming inputs and decides in real-time when to slice the input frame.
The inputs in the queue have a specific Time To Live (TTL), after which they are discarded.
When the number of inputs in the queue surpasses a threshold (T), the system begins the procedure of defining the frame.
It waits for a time equal to half of the TTL, then gathers all the inputs in the queue, passes them to the aggregator component, and clears the input queue. 
It is crucial to correctly estimate TTL, as it creates a trade-off between the responsiveness of the system and its accuracy, as illustrated in Figure~\ref{fig:frames-dynamic}.

When the Dynamic Input Frame is used together with the Continuous Reliability System (\S~\ref{subsec:input_aggregation}), instead of using the number of inputs to compare to the threshold (T) we calculate a weighted sum using players' reliabilities, so that reliable players have more control over when the input frame is issued.

\subsection{Inputs Aggregator}
\label{subsec:input_aggregation}

The second component of the Crowdsourcing Controller gets all the inputs within the input frame and returns a single aggregated input.
We consider two approaches to aggregate inputs: (1) Majority Vote (MV); and (2) our Continuous Reliability System.




MV simply selects the most popular input among players.
Ties are arbitrarily broken with a predefined order of preferred inputs.
Note that no or missing input also counts as a valid input, because sometimes the correct action is not issuing any commands at all.

The Continuous Reliability System uses weighted MV, where the votes are weighted according to the reliability score of each player, i.e., an estimate of the ability of each player to issue correct inputs.
Next, we describe how to compute and update reliabilities in real-time.



Inspired by the idea of Collective Intelligence, the Continuous Reliability System  rewards players that issue the same input as the majority and punishes those that vote for unpopular options.
The system starts by aggregating inputs with MV, i.e., uniform reliabilities.
Every time the system receives an input frame, the reliability scores are updated proportionally to how popular the players' votes are.
For example, when $95\%$ of the players vote for input A and $5\%$ vote for input B, the reliabilities of the players that voted for option B will drop significantly, compared to a situation where a split between A and B is $60\%$ to $40\%$.
After a certain number of iterations, reliability scores might change the final output: the aggregated input is not the most popular one, but a less popular option can be chosen because of the high reliability of the players who issued that command.
This system is completely decoupled from the game itself, it does not need to be aware of whether an input was \textit{good} or \textit{bad} based on the game state, which makes it easier to apply it to other scenarios in the future. 



Next we formally describe the Continuous Reliability algorithm (Algorithm~\ref{fig:rel-alg}).
Let $\mathcal{I} = \{0,1, \dots K-1\}$ be the list of all possible inputs issued by players and $N_p$ be the number of players.
A single input frame $\mathcal{F}$ is an $N_p$-dimensional vector with values from $\mathcal{I}$, i.e., the votes of the players within the time frame.
$R_j$ is the reliability score of player $j$ and $R$ is the $N_p$-dimensional vector containing all current reliability scores.

The Continuous Reliability algorithm is recurrent.
At the start, each player is assigned the same reliability score of $1$, then at each iteration we compute $\Delta R$, i.e., the vector of reliability changes that is added to update reliability scores.

Consider a fixed iteration with input frame $\mathcal{F}$. 
We define $\mathcal{V} = \{ v_0, v_1, \dots v_{K-1} \}$ as a $K$-dimensional vector counting the number of votes for each input in $\mathcal{F}$.
We say that an input $j$ is \textit{viable} when it receives ``enough'' votes, i.e., $v_j \geq \gamma*v_{max}$, where $\gamma$ is an \textit{agreement threshold} and $v_{max} = \text{max}(\mathcal{V})$, i.e., the maximum number of votes achieved by a single input. 
This handles cases when the votes are evenly split between more than one inputs: we do not want to decide that only one input is correct, but we will consider all viable votes when updating player reliabilities.

If all inputs in $\mathcal{F}$ are viable we do not change reliabilities, because all players selected a potentially correct input. 
If there are non-viable inputs, we proceed to compute $\Delta R$.

First we treat non-viable inputs in $\mathcal{F}$, intuitively they should affect $\Delta R$ in a proportional way: the less popular an input, the lower the player reliability who issued that. 
We define this as follows: 
\begin{equation}
\Delta R_{{j}_{nv}} = - \frac{v_{max}}{v_{\mathcal{F}_j}} * \delta
\label{eq:rel_nv}
\end{equation}
where $j$ is a player who issued a non-viable input; $v_{\mathcal{F}_j}$ is the number of votes received by the input issued by player $j$; and $\delta\geq0$ is the \textit{reliability step coefficient}, which controls how much a viable or non-viable input affects the reliability score. The higher $\delta$, the faster the system will adjust the reliabilities.
The current reliability of player $j$, $R_j$, is updated by adding (subtracting) the quantity in Equation~\eqref{eq:rel_nv}.


Note that the reliability scores might fall below $0$, which means that unreliable players will in fact decrease the number of votes for the input they choose. This mechanism can be effective when dealing with players that issue wrong inputs on purpose, as they still unknowingly help the system in selecting the correct input.
On the other side, a limitation of this mechanism arises when the same player keeps on issuing wrong inputs, e.g., inactivity, network lag, etc.
If the player resumes the game and starts issuing correct inputs, his/her very low reliability will disrupt the voting, since weighted MV will heavily favor options not chosen by this player.
Therefore, we introduce a \textit{lower bound} $\omega\geq0$, which clamps the change of the reliability if it brings the total player reliability below $-\omega$:
\begin{equation*}
     \Delta R_{{j}_{nv}} = 
\begin{cases}
    \frac{v_{max}}{v_{\mathcal{F}_j}} * \delta ,& \text{if } R_{j} + \Delta R_{{j}_{nv}} \geq - \omega\\
    \omega - R_{j},              & \text{otherwise}
\end{cases}
\end{equation*}
where $R_j$ is the current reliability score for player $j$.
The opposite problem is not a concern for the upper bound. For players to have a high reliability score, they not only have to be consistent in issuing the input, but also other players have to vote similarly to them. This makes a situation where one of the players has a significantly larger reliability score than others very unlikely.

Then, we treat viable inputs in $\mathcal{F}$. Again they should affect $\Delta R$ in a proportional way: the more popular an input, the higher the player reliability who issued that.  
We do this as follows:
\begin{equation}
\Delta R_{{j}_{v}} = v_{\mathcal{F}_j} * \frac{\sum_{k\in nv} \Delta R_k}{\sum_{k\in v} v_{\mathcal{F}_k}}
\end{equation}
where the first factor $v_{\mathcal{F}_j}$ is the number of votes received by the input issued by player $j$ and the second factor is a normalization factor.
The normalization factor is computed as the ratio of the sum $\Delta R$ for non-viable inputs and the sum of $v_{\mathcal{F}_k}$ for the viable inputs. 
The normalization factor is needed to guarantee that the sum of reliabilities remains constant and equal to the number of players, which is needed by weighted MV.




\begin{algorithm}[t]
\SetAlgoLined
\textbf{Input: } Players input frame $\mathcal{F}$, current reliabilities $R = [R_0, \dots R_{N_p - 1}]$   \\
\KwResult{Vector of reliabilities updates $\Delta R = [\Delta R_0, \dots \Delta R_{N_p - 1}]$}

 $\Delta R = [0, 0, \dots 0]_{N_p}$\\
 $v_{\mathcal{F}_j} = \sum_k (\mathcal{F}_j == \mathcal{F}_k) $  \tcp{\small{Number of votes for the input issued by player j}}
 \If{$\forall j \in viable(v_{\mathcal{F}_j})$}{
    \KwRet{$\Delta R$}
  }
   \tcp{\small{First pass - non-viable votes}}
 \For{not $viable(v_{\mathcal{F}_j}) \in \mathcal{V}$}
 { 
        $\Delta R_j = - \frac{v_{max}}{v_{\mathcal{F}_j}} * \delta$ \\
      \If{$R_j + \Delta R_j \leq -\omega$}{
       $\Delta R_j = \omega - R_j$
  }
 }
 $norm = \frac{\sum_{k\in nv} \Delta R_k}{\sum_{k\in v} v_{\mathcal{F}_k}}$ \tcp{\small{Viable votes normalization factor}}
    \tcp{\small{Second pass - viable votes}}
  \For{$viable(v_{\mathcal{F}_j}) \in \mathcal{V}$}{
    $\Delta R_j = v_{\mathcal{F}_j} * norm $ 
}
\KwRet{$\Delta R$}
 \caption{Calculating the reliability updates}
 \label{fig:rel-alg}
\end{algorithm}

\section{Experimental Evaluation}
\label{sec:evaluation-em}

Next we evaluate the performance of the Crowdsourcing Controller, i.e., input frames and input aggregators.




\subsection{Experimental Setup}

The Crowdsourcing Controller depends on a number of parameters.
For static input frames we need to set only the frame length.
For dynamic input frames we need to set the TTL and the frame threshold T, i.e., the number of inputs from different users needed to determine the centre of the dynamic time window (see Figure~\ref{fig:frames-dynamic}).
We set TTL to $300$ms, so on average the time between issuing an input and seeing a response on the screen is around $150$ms. Threshold $T = 0.5$ was set so that half of the inputs marked the centre of the dynamic frame.
For the continuous reliability system, we need to estimate the agreement threshold $\gamma \in [0, 1]$, i.e., the proportion of votes needed to define a viable input, and the reliability step coefficient $\delta \in [0, 1]$, which determines the speed in updating reliabilities.
The parameters have been fine-tuned for offline and online experiments during development and used consistently to compare the system under the same circumstances. The values were chosen so that the learning process of the system can stabilize in few minutes, as longer sessions could cause players' fatigue. The values used were $\gamma = 0.8$ (offline), $\gamma = 0.6$ (online) and  $\delta = 0.05$. The lower bound $\omega$ is set to $0.5$.

For the input frame component, we consider the simple static input frame approach as our baseline opposed to our dynamic input frame approach.
For the inputs aggregator component we consider MV as baseline opposed to our Continuous Reliability System.
In the experiments we consider $4$ different system configurations: (1) static frame with Continuous Reliability System; (2) Dynamic Input Frame with Continuous Reliability System; (3) Static frame with MV; (4) Dynamic Input Frame with MV.

For the evaluation, there is no available ground-truth, i.e., single correct sequence of inputs, that we can use to evaluate the Crowdsourcing Controller.
This is because the game is dynamic and there is more than one way of beating obstacles, e.g., some obstacles can be beaten with both short and long jumps.
Therefore we use the game score as evaluation measure.
The game score represents how long it takes for the players to lose the game, by issuing the wrong type of input or providing it at the incorrect time.
The higher the game score the better.

Additionally we consider EM~\cite{Dawid79maximumlikelihood} as our reference or ``ground-truth''.
Due to its iterative definition, EM can not be run online, therefore it is run offline when all data are gathered.
Then we compute the agreement between one of the $4$ configurations and EM, i.e., the accuracy with the inputs issued by a configuration and the inputs predicted by EM as ground-truth. 
The higher the agreement the better.

\subsection{Offline Experiment with Synthetic Players}
\label{subsection:offline_experiments}


In this experimental scenario we evaluate the Crowdsourcing Controller in an artificial setting with synthetic players.
Players are replaced with AIs designed to represent humans of different skills.
The \textit{Perfect AI} always issues a correct input at each time-stamp during the game.
All other AIs are defined by adding noise to the perfect AI behaviour.
This is done with $3$ parameters representing: noise, noise shift and chance of error.
We generated a total of $14$ different AIs: $1$ Perfect, $3$ Good, $4$ Bad, $1$ Shifted (-), $2$ Shifted (+), $1$ Majorly Shifted (-), $1$ Confused, and $1$ Random.
Each AI was individually tested 3 times and their mean scores are in Table~\ref{tab:ais-solo-scores}.



\begin{table}[tb]
\centering
\caption{AIs individual game scores}
\label{tab:ais-solo-scores}
\begin{tabular}{|l|r|r|}
\hline
\textbf{Name}       & \multicolumn{1}{l|}{\textbf{Average Score}} & \multicolumn{1}{l|}{\textbf{Max Score}} \\ \hline
Perfect             & 1182                                                                & 1353                                                            \\ \hline
Good                & 250                                                                 & 488                                                             \\ \hline
Bad                 & 63                                                                  & 66                                                              \\ \hline
Confused            & 23                                                                  & 28                                                              \\ \hline
Random              & 48                                                                  & 64                                                              \\ \hline
Shifted (-)         & 164                                                                 & 398                                                             \\ \hline
Shifted (+)         & 27                                                                  & 30                                                              \\ \hline
Majorly Shifted (-) & 98                                                                  & 175                                                             \\ \hline
\end{tabular}
\end{table}




\begin{figure}[tb]
  \centering
  \includegraphics[width=8cm]{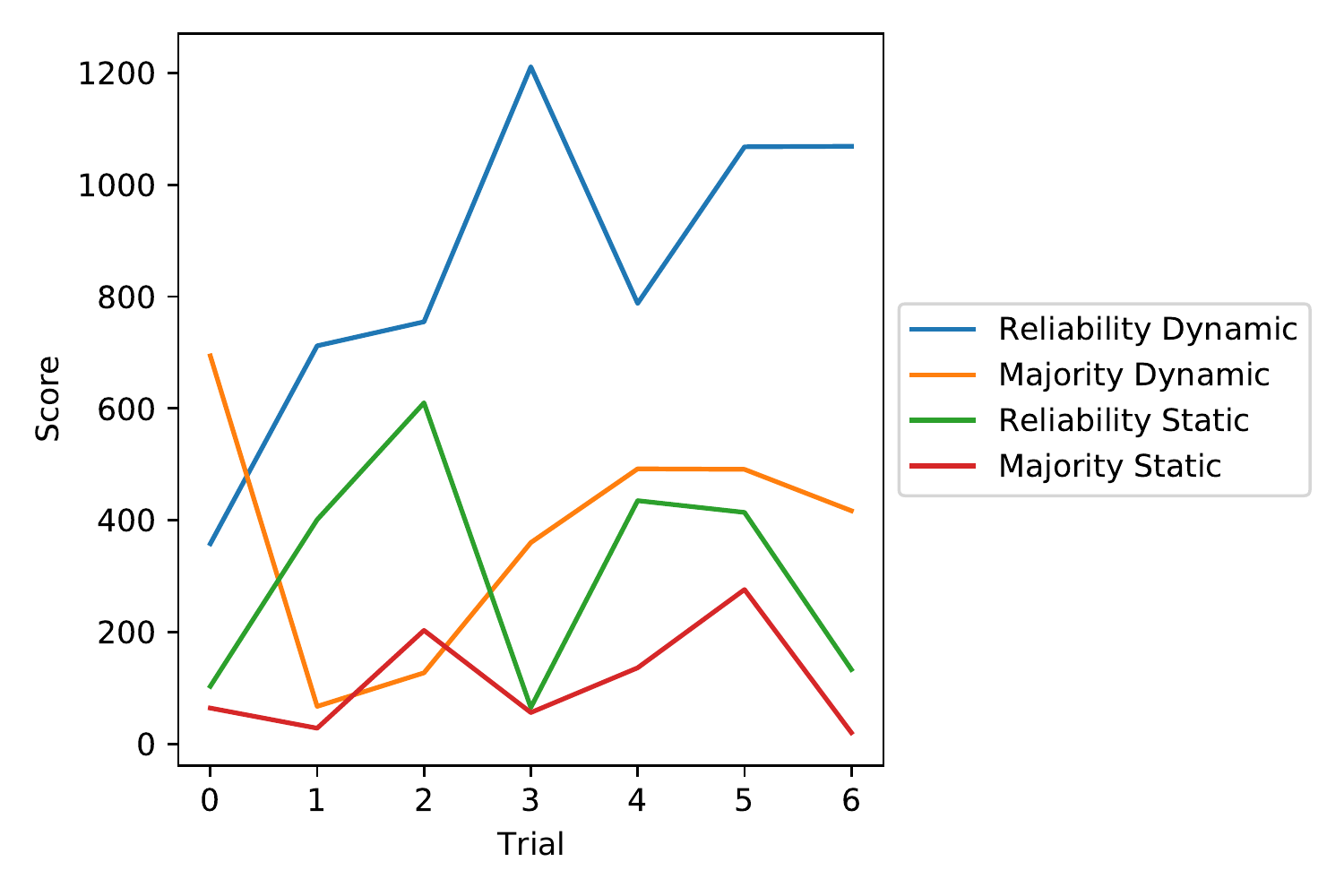}
  \caption{14 AIs scores with different system configurations}
    \label{fig:big-ai-scores}
\end{figure}


\begin{table}[tb]
\centering
\caption{14 AIs experiment results}
\label{tab:big-experiment-results}
\begin{tabular}{|l|c|c|c|}
\hline
\multicolumn{1}{|c|}{\textbf{Name}} &
  \textbf{\begin{tabular}[c]{@{}c@{}}Aggrement\\ w/ EM predictions\end{tabular}} &
 \textbf{\begin{tabular}[c]{@{}c@{}} Mean\\ score\end{tabular}} &
  \textbf{\begin{tabular}[c]{@{}c@{}} Max\\ score\end{tabular}} \\ \hline
Dynamic Reliability & 0.9937 & 852 & 1211 \\ \hline
Dynamic Majority  & 0.9915 & 474 & 981  \\ \hline
Static Reliability & 0.9553 & 309 & 610  \\ \hline
Static Majority & 0.9714 & 94  & 277  \\ \hline
\end{tabular}
\end{table}

The comparison of scores achieved by different system configurations across $6$ consecutive trials is presented in Figure~\ref{fig:big-ai-scores} and in Table~\ref{tab:big-experiment-results}.
We stopped after $6$ trials since we reached convergence of the reliability scores, i.e., after trial $6$ there was little change in reliability scores.

From the results we can see that the Dynamic Input Frame configurations outperform the static ones, both in terms of game score and agreement with EM.
Moreover, the configurations with the Continuous Reliability system outperformed MV, both in terms of scores and similarity with respect to EM. 
For the Dynamic Input Frame with Continuous Reliability System (the best configuration), the agreement with EM reached a value of $99.37\%$ compared to $94.40\%$ if the system is replaced with majority voting ($5.26\%$ relative improvement). 
In Figure~\ref{fig:big-ai-scores}, the Dynamic Input Frame with Continuous Reliability System outperforms all other configurations already after $1$ trial.
This means that the system quickly and correctly learns which players are more reliable. 

\begin{figure}[tb]
  \centering
  \includegraphics[width=8cm]{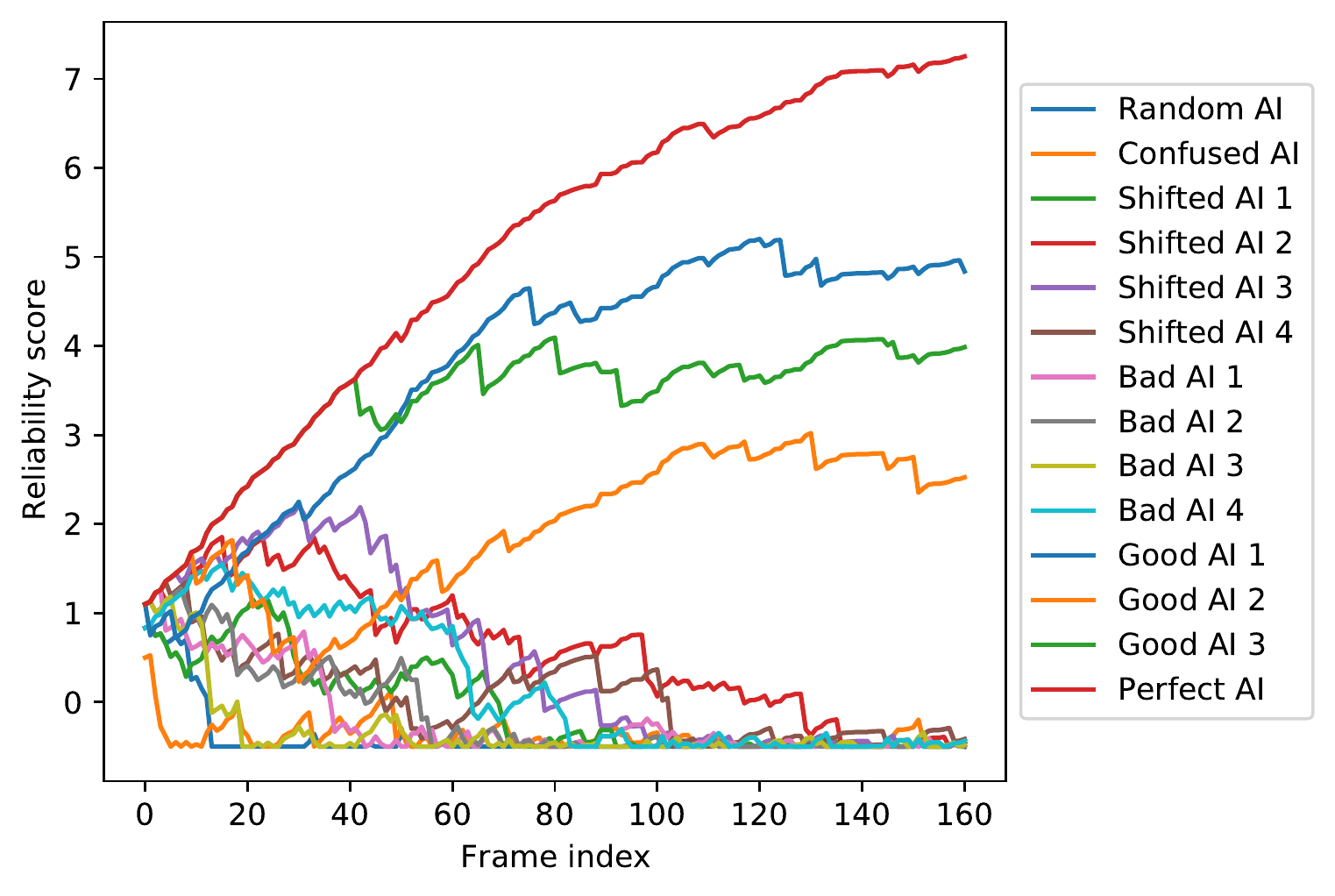}
  \caption{14 AIs reliability training with the Dynamic Input Frame}
    \label{fig:ai-reliability}
\end{figure}

Figure~\ref{fig:ai-reliability} shows the process of the reliability training with the Dynamic Input Frame configuration.
The reliability scores ($y$-axes) are plotted for each of the analyzed input frames ($x$-axis). 
The process is chaotic at the beginning but stabilizes after circa $100$ frames.
After $100$ frames, the estimated reliabilities correlate with the AIs skills: the top reliability score (red line) belongs to the Perfect AI.
The inputs issued by the perfect AI will be weighted as $7$ players.
The following $3$ reliabilities scores belong to the $3$ Good AIs.
At the end of the experiment, the Perfect AI with the help of just one of the Good AIs can outvote the rest $12$ AIs combined.
Indeed, the top $4$ AIs alone have an impact of over $18$ players. 
This exceeds the total number of players ($14$) because the rest of the reliabilities are less than $0$.
We can also observe the effect of the lower bound $\omega$, which prevents bad players from disrupting the weighted voting system.

\subsection{Online Experiment with Real Players}
\label{subsec:user-studies}




In this experimental scenario, we evaluate the proposed Crowdsourcing Controller with real users.
We tested only $3$ configurations out of $4$: Dynamic Input Frame with MV, Dynamic Input Frame with Continuous Reliability System, and static frame with Continuous Reliability System.
We did not test the static frame with MV because: (1) the experiment with $4$ configurations was too long and players' fatigue could affect the quality of the collected data; (2) the offline experiments show that this is the worst performing configuration (see \S~\ref{subsection:offline_experiments})

We enrolled a total of $9$ players, all of them were university students with a scientific background.
Before starting the experiment, participants were trained by playing individually for $3$ minutes. 
Afterward they had $3$ trials to achieve their highest scores. 
Then we Started the multiplayer phase. 
Each configuration was played for $3$ minutes for training, followed by $5$ trials to beat the highest score. 
The whole process was repeated $2$ times to collect more data. Exact user instructions can be found in our code repository. 
During the experiments the reliability scores were not shown to the players. This was done to avoid the possibility of the players feeling discouraged due to low impact of their inputs. 
There was a short break between the sets of configurations to give the players a chance to rest.


\begin{figure}[tb]
  \centering
  \includegraphics[width=8cm]{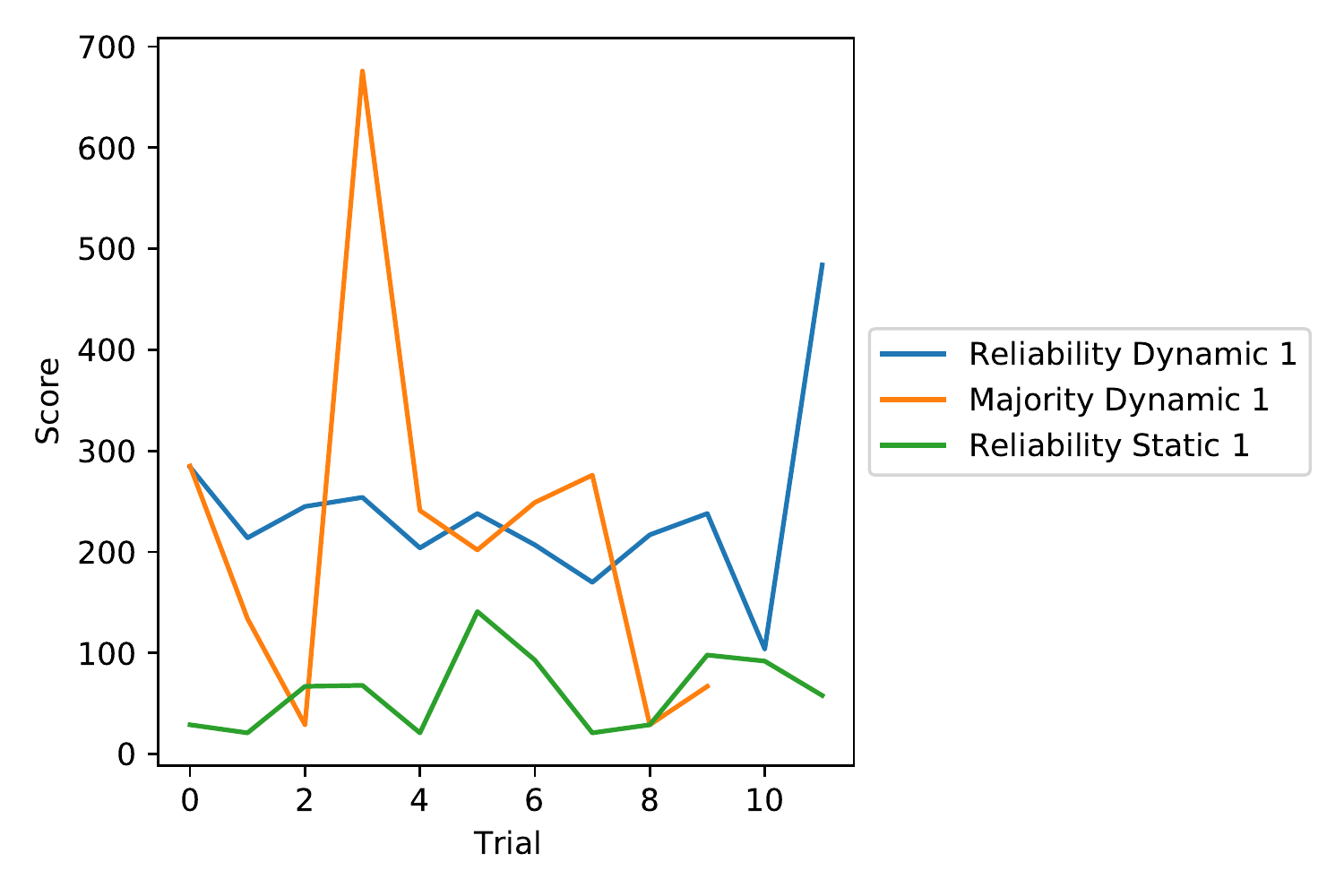}
  \caption{User scores during first set of trials}
    \label{fig:user-scores-01}
\end{figure}

\begin{table}[tb]
\centering
\caption{User studies result summary}
\label{tab:user-experiment-results}
\begin{tabular}{|l |c |c|c|}
\hline
\multicolumn{1}{|c|}{\textbf{Name}} &
  \textbf{\begin{tabular}[c]{@{}c@{}}Aggrement\\ w/ EM predictions\end{tabular}} &
 \textbf{\begin{tabular}[c]{@{}c@{}} Mean\\ score\end{tabular}} &
  \textbf{\begin{tabular}[c]{@{}c@{}} High\\ score\end{tabular}} \\ \hline
Dynamic Reliability & 0.9619 & 255 & 541 \\ \hline
Dynamic Majority   & 1.00  & 194 & 516 \\ \hline
Static Reliability & 0.9108 & 76  & 434 \\ \hline
\end{tabular}
\end{table}

The game scores achieved by the players during the first set of trials (both training and game) are shown in Figure~\ref{fig:user-scores-01}.
The results for the second trial are comparable.
Table~\ref{tab:user-experiment-results} reports the results aggregated over all trials.

Online experiments corroborate the observations from offline experiments with respect to input frames.
The Dynamic Input frame configurations outperform the static input frame configuration (green line in Figure~\ref{fig:user-scores-01}). 
This is true both in terms of game score and agreement with EM.
%
Second, in terms of game score, the Dynamic Input Frame with Continuous Reliability outperforms the other $2$ configurations.

\begin{figure}[tb]
  \centering
  \includegraphics[width=8cm]{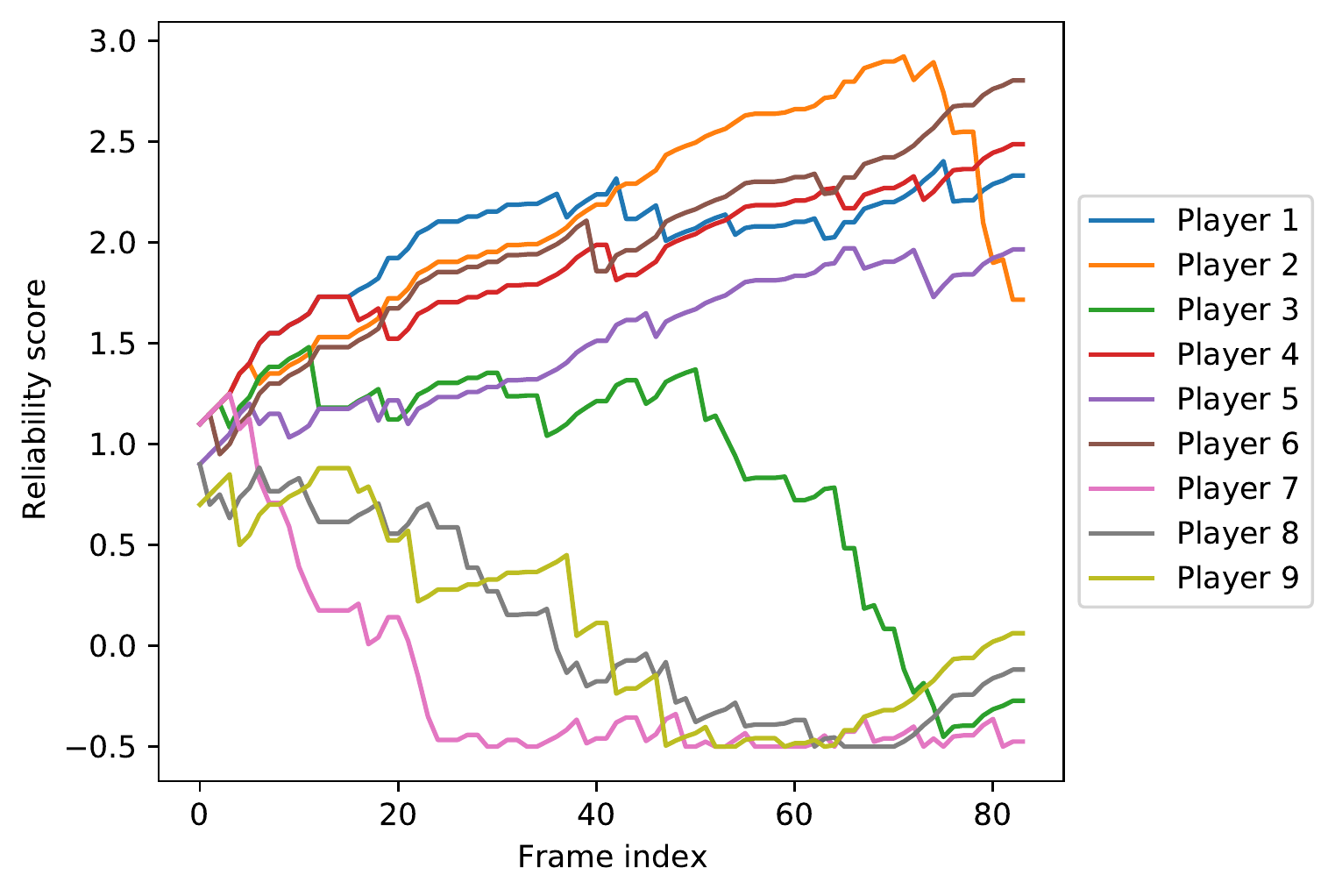}
  \caption{Player reliability training in one of the sessions with the Dynamic Input Frame}
    \label{fig:player-reliabilities}
\end{figure}

Online experiments differ from offline experiment with respect to the agreement with EM.
We can observe that using a real-time MV system with dynamic frames yields the same results as using an offline-trained EM.
This suggests that the majority of the players are similarly skilled, thus using a weighted MV instead of the uniform MV does not make a significant difference.
We further analyzed the agreement among the players which showed that a particular majority (specifically $5$ players out of $9$) agreed with each other on issuing the same input over $80\%$ of the times.
This can be seen also in Figure~\ref{fig:player-reliabilities}, where the reliability scores of the top $5$ players are clearly separated from the rest of the group.

Even if all the players are similarly skilled, the better performance of the Continuous Reliability system in terms of game scores might come from the fact that fewer players with stronger influence  more quickly surpass the needed threshold of votes required by the Dynamic Input Frame, which helps with beating obstacles that require fast decisions.


At the end of the user studies, participants filled a short survey. Among different observations, over half of the participants mentioned, in different words, that they focused more during multiplayer than individual sessions, which is aligned with previous results about gamification approaches to encourage crowd workers (see Section~\ref{subsec:crowdsourcing}). 
Some participants mentioned the social aspect of playing together, saying that winning and losing together was more \textit{fun}.
This suggests that the system could be used as a gameplay mechanic in EMGs. Some players, however, reported feeling frustrated when they lost despite seemingly issuing correct inputs, sometimes blaming it on the network delay or software errors. This implies that players could benefit from even more visual feedback regarding the decision making of the Crowdsourcing Controller.

\subsection{Discussion} 
\label{subsec:discussion}

Next, we summarize the main conclusions and observations from offline and online experiments.
The Dynamic Input Frame proves to be more efficient than a static approach proposed earlier in the literature. 
It is able to achieve much higher scores thanks to more precise predictions of the correct time that the input should be issued at.

The Continuous Reliability System increases the game score to a large extent for offline experiment, but to a smaller extent for online experiments.
The reason for this is the observed lack of differences in the skills of real players. Although a small number of them did in fact perform worse than others, the majority usually agreed on the input type and time. In this situation, it is expected that MV gives similar results as the Continuous Reliability System. Moreover, analyzing players' inputs proves that participants usually agreed on the type of the input, with the differences mostly showing in the precise time of issuing it. The reason for this comes from the nature of the game that was used to test the system. The real challenge was in fact issuing the input at the right time, since choosing the right input was relatively simple as different obstacles were easily distinguishable. 

Both user surveys and observations during the study showed a significant social aspect of playing the game together, where synchronization seemed like one of the goals that the players wanted to achieve, knowing it would raise their chances of getting high scores.

\section{Conclusions and Future Work}
\label{subsec:conclusion}

In this paper we present a prototype of an online video game, where multiple players control simultaneously the same character.
We propose the Crowdsourcing Controller, i.e., a system that: (1) groups real-time inputs from multiple players into time frames; (2) aggregates the inputs within a frame and returns a single output command.
We propose the Dynamic Input Frame method, which adjusts the frame length dynamically based on the inputs received.
To aggregate the inputs, we propose the Continuous Reliability System, which computes a weighted MV by estimating players' reliabilities.
Offline and online experiments show that the Dynamic Input Frame is better than state-of-the-art static frames and that  the Continuous Reliability System leads to higher game scores.




As future work we plan to test the proposed crowdsourcing system with a larger and more diverse set of players.
Moreover, we will explore how the proposed systems cope in different settings,  for example scenarios featuring more agency and creativity on the players' side.
Finally, we want to inspect the possibilities of automatically optimizing some of the system parameters that were arbitrarily chosen, for example the TTL of the Dynamic Input Frame.

\vspace{12pt}

\end{document}